\documentclass[twocolumn,prb,aps,showpacs]{revtex4-1}
\usepackage{graphicx}
\usepackage{color}
\usepackage{dcolumn}
\usepackage{amsmath}
\usepackage{units}

\newlength{\figwidth} 
 \newlength{\figwidthb} %
\newcommand{\NIO}{Na$_2$IrO$_3$ } 
\newcommand{\NIOns}{Na$_2$IrO$_3$}

\newcommand{\SIOns}{Sr$_2$IrO$_4$} 

\begin{document}

\title{Direct Evidence for Dominant Bond-directional Interactions in a Honeycomb Lattice Iridate \NIO} 

\author{Sae Hwan Chun$^1$, Jong-Woo Kim$^2$, Jungho Kim$^2$, H. Zheng$^1$, Constantinos C. Stoumpos$^1$, C. D. Malliakas$^1$, J. F. Mitchell$^1$, Kavita Mehlawat$^3$, Yogesh Singh$^3$, Y. Choi$^2$, T. Gog$^2$, A. Al-Zein$^4$, M. Moretti Sala$^4$, M. Krisch$^4$, J. Chaloupka$^5$, G. Jackeli$^{6,7}$, G. Khaliullin$^6$ and B. J. Kim$^6$} 

\address{$^1$Materials Science Division, Argonne National Laboratory, Argonne, IL 60439, USA\\$^2$Advanced Photon Source, Argonne National Laboratory, Argonne, IL 60439, USA\\$^3$Indian Institute of Science Education and Research (IISER) Mohali, Knowledge City, Sector 81, Mohali 140306, India\\$^4$European Synchrotron Radiation Facility, BP 220, F-38043 Grenoble Cedex, France\\$^5$Central European Institute of Technology, Masaryk University, Kotlářská 2, 61137 Brno, Czech Republic\\$^6$Max Planck Institute for Solid State Research, Heisenbergstra\ss e 1, D-70569 Stuttgart, Germany\\$^7$Institute for Functional Matter and Quantum Technologies, University of Stuttgart, Pfaffenwaldring 57, D-70569 Stuttgart, Germany
	} 
	
\date{\today}




\maketitle
\noindent
{\bf Heisenberg interactions are ubiquitous in magnetic materials and have been prevailing in modeling and designing quantum magnets. Bond-directional interactions\cite{vanVleck,Khaliullin05,Jackeli09} offer a novel alternative to Heisenberg exchange and provide the building blocks of the Kitaev model\cite{Kitaev}, which has a quantum spin liquid (QSL) as its exact ground state. Honeycomb iridates, A$_2$IrO$_3$ (A=Na,Li), offer potential realizations of the Kitaev model, and their reported magnetic behaviors may be interpreted within the Kitaev framework. However, the extent of their relevance to the Kitaev model remains unclear, as evidence for bond-directional interactions remains indirect or conjectural. Here, we present direct evidence for dominant bond-directional interactions in antiferromagnetic \NIO and show that they lead to strong magnetic frustration. Diffuse magnetic x-ray scattering reveals broken spin-rotational symmetry even above T$_{\textrm N}$, with the three spin components exhibiting nano-scale correlations along distinct crystallographic directions. This spin-space and real-space entanglement directly manifests the bond-directional interactions, provides the missing link to Kitaev physics in honeycomb iridates, and establishes a new design strategy toward frustrated magnetism.   }

\noindent
Iridium (IV) ions with pseudospin-1/2 moments form in \NIO a quasi-two-dimensional (2D) honeycomb network, which is sandwiched between two layers of oxygen ions that frame edge-shared octahedra around the magnetic ions and mediate superexchange interactions between neighboring pseudospins (Fig.~1a). Owing to the particular spin-orbital structure of the pseudospin\cite{Kim08,Kim09}, the isotropic part of the magnetic interaction is strongly suppressed in the 90$^\circ$ bonding geometry of the edge-shared octahedra\cite{Khaliullin05,Jackeli09}, thereby allowing otherwise subdominant bond-dependent anisotropic interactions to play the main role and manifest themselves at the forefront of magnetism. This bonding geometry, common to many transition-metal oxides, in combination with the pseudospin that arises from strong spin-orbit coupling gives rise to an entirely new class of magnetism beyond the traditional paradigm of Heisenberg magnets. On a honeycomb lattice, for instance, the leading anisotropic interactions take the form of the Kitaev model\cite{Jackeli09}, which is a rare example of exactly solvable models with nontrivial properties such as Majorana fermions and non-abelian statistics, and with potential links to quantum computing\cite{Kitaev}.

\begin{figure*}
\centerline{\includegraphics[width=1.55\columnwidth,angle=0]{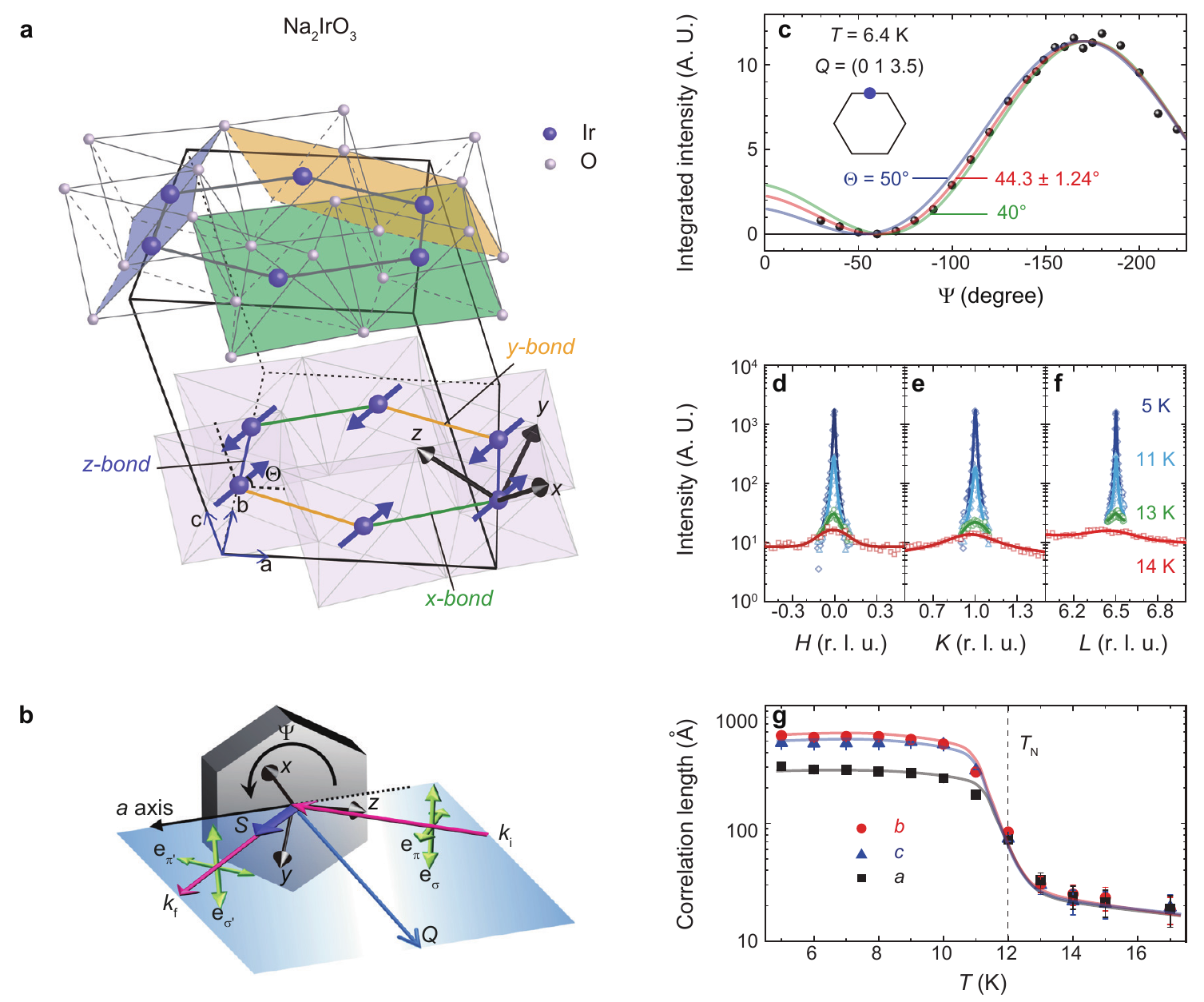}}
\caption{{\bf Magnetic easy axis and temperature dependence of the zig-zag order.} (a) Honeycomb layers of Ir$^{4+}$ in the monoclinic Bravais lattice. Green, yellow, and blue planes show Ir$_2$O$_6$ plaquettes normal to local $x$, $y$, and $z$ axes (black arrows), respectively, which point along three nearest neighbor Ir-O bonds in an octahedron. Ir-Ir bonds are labeled following the plaquettes they belong to. Na$^+$ is not shown for clarity. Blue arrows show the spins in the static zig-zag order propagating along the $b$ direction. Spins are antiparallel between the layers (not shown). (b) Illustration of the scattering geometry. Shown in blue is the scattering plane defined by the incident (${\mathbf k_i}$) and outgoing (${\mathbf k_f}$) wavevector (red arrows). Green arrows show the x-ray polarizations. The azimuth $\Psi$ is defined as the angle between the $a$ axis and the scattering plane. (c) $\Psi$-dependence of the magnetic Bragg peak (blue filled circle) intensity at (0 1 3.5) measured in the $\sigma$-$\pi'$ channel. Black hexagon is the Brillouin zone of the honeycomb net. Red solid line shows the best fitting result to the data with $\Theta = 44.3^{\circ}$. Green and blue lines shows the calculated $\Psi$ dependence for $\Theta = 40^{\circ}$ and $50^{\circ}$, respectively. (d-f) H, K, and L scans of the magnetic Bragg peak at (0 1 6.5) for selected temperatures. (g) Temperature dependence of the correlation lengths along the $a$, $b$, and $c$ axes from Gaussian fitting to the scans. Error bars represent the standard deviation in the fitting procedure. The solid lines are guides to eye.}\label{fig:fig1}
\end{figure*}

Realization of the Kitaev model is now being intensively sought out in a growing number of materials\cite{Takayama,Modic,Kimchi14,Hermanns,Lee14,Plumb14,Luo13}, including 3D extensions of the honeycomb Li$_2$IrO$_3$, dubbed ``hyper-honeycomb''\cite{Takayama} and ``harmonic-honeycomb''\cite{Modic}, and 4d transition-metal analogs such as RuCl$_3$\cite{Plumb14} and Li$_2$RhO$_3$\cite{Luo13}. Although most of these are known to magnetically order at low temperature, they exhibit a rich array of magnetic structures including zig-zag\cite{Liu11,Choi12,Ye12}, spiral\cite{Reuther14}, and other more complex non-coplanar structures\cite{Biffin14,Biffin142} that are predicted to occur in the vicinity of the Kitaev QSL phase\cite{Chaloupka10,Chaloupka13,Lee,Rau14}, which hosts many degenerate ground states frustrated by three bond-directional Ising-type anisotropies. All of these magnetic orders are captured in an extended version of the Kitaev model written as
\begin{align}
	H &= \sum_{\langle ij\rangle_\gamma } 
	[K S_i^\gamma S_j^\gamma + J {\mathbf S_i} \cdot {\mathbf S_j}+\Gamma(S_i^\alpha S_j^\beta+S_i^\beta S_j^\alpha) ],
\end{align}
\noindent
which includes, in addition to the Kitaev term $K$, the Heisenberg exchange $J$ that may be incompletely suppressed in the superexchange process and/or arise from a direct exchange process\cite{Chaloupka13}, and the symmetric off-diagonal exchange term $\Gamma$, which is symmetry-allowed even in the absence of lattice distortions\cite{Rau14,Katukuri14,Yamaji14}. This ``minimal'' Hamiltonian couples pseudospins $\mathbf{S}$ (hereafter referred to as `spin') only on nearest-neighbor bonds $\langle ij\rangle$, neglecting further-neighbor couplings, which may be non-negligible. The bond-directional nature of the $K$ and $\Gamma$ terms is reflected in the spin components  [$\alpha\neq\beta\neq\gamma\in(x,y,z)$] which they couple for a given bond (
$\gamma\in x-,y-,z-$bonds)(Fig.~1a). For example, the $K$ term couples only the spin component normal to the Ir$_2$O$_6$ plaquette containing the particular bond. Despite these extra terms that may account for finite-temperature magnetic orders in the candidate materials, the fact that the Kitaev QSL phase has a finite window of stability against these perturbations\cite{Chaloupka10,Yamaji14} calls for investigation of competing phases and a vigorous search for the Kitaev QSL phase. 

Although the notion of magnetic frustration induced by competing bond-directional interactions is compelling, it remains a theoretical construct without an existential proof for such interactions in a real-world material. Moreover, theories for iridium compounds based on itinerant electrons suggest alternative pictures\cite{Shitade09,Mazin12,KimC12}. In principle, measurement of the dynamical structure factor through inelastic neutron scattering (INS) or resonant inelastic x-ray scattering (RIXS) provides the most direct access to the Hamiltonian describing the magnetic interactions. However, a fully momentum and energy resolved dynamical structure factor thus far remains elusive for any of the candidate materials; RIXS suffers from insufficient energy resolution\cite{Gretarsson13} and INS is currently limited by unavailability of large-volume single crystals\cite{Choi12}. In this Letter, we take a new approach using diffuse magnetic x-ray scattering to provide direct evidence for predominant bond-directional interactions in \NIO via the measurement of equal-time correlations of spin components above the ordering temperature (T$_{\textrm N}=12-$15 K, See Fig.~S1).

%
%
\begin{figure*}
\hspace*{-0.2cm}\vspace*{-0.1cm}\centerline{\includegraphics[width=1.55\columnwidth,angle=0]{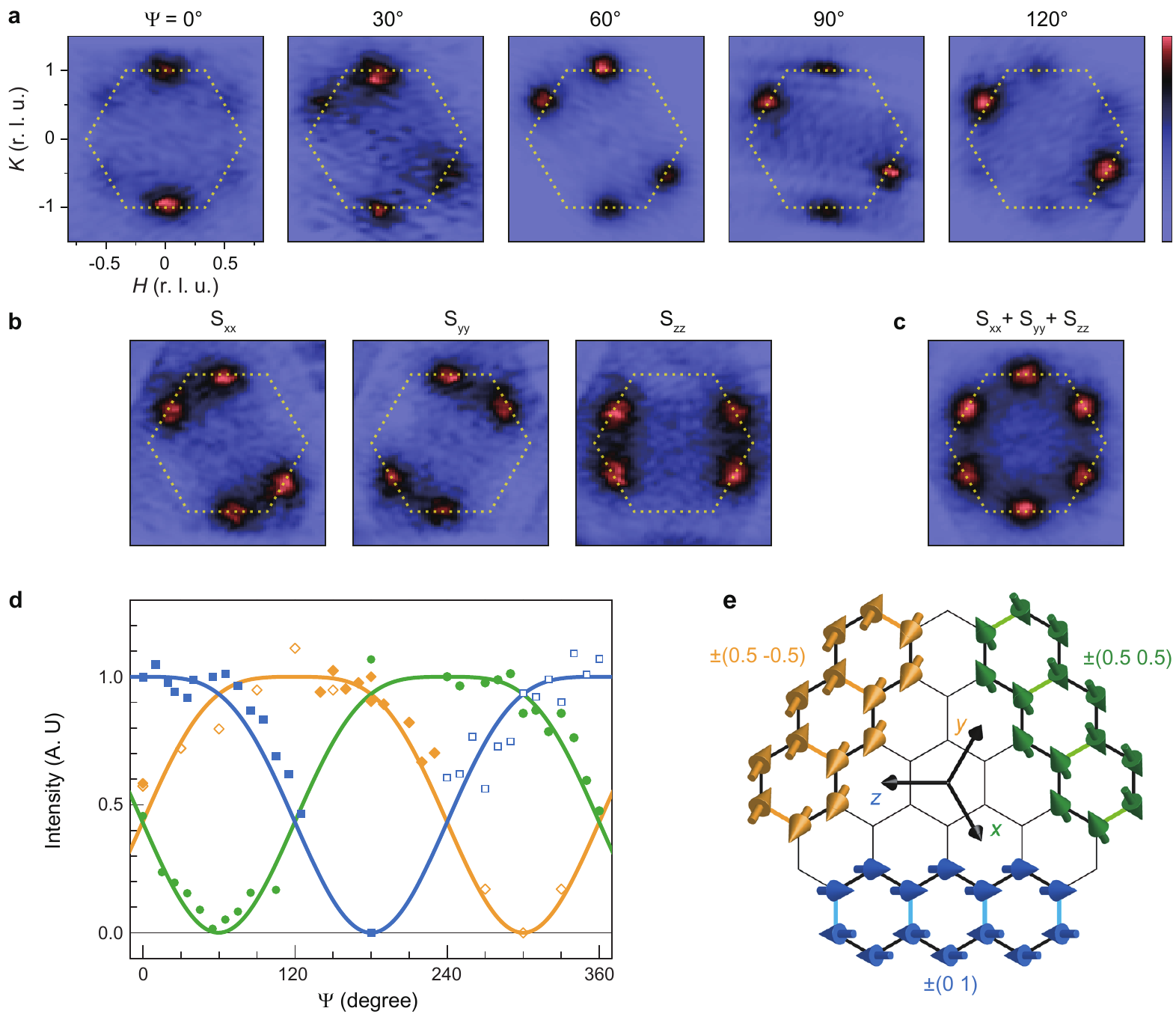}}
%
\caption{{\bf Diffuse magnetic x-ray scattering intensities above T$_{\textrm N}$.} (a) Intensity plots in the HK plane measured at T = 17 K for selected azimuth angles summing $\pi$-$\sigma'$ and $\pi$-$\pi'$ channels, sensitive to spin components parallel to ${\mathbf k_i}$ and perpendicular to the scattering plane, respectively. For example, $\Psi=0^\circ$ measures sum of correlations $S_{xx}$ and $S_{yy}$. The dashed hexagon indicates the first Brillouin zone of the honeycomb net. (b) Spin-component-resolved equal-time correlations extracted from (a). (c) Spin-component-integrated equal-time correlations extracted from (a). Peaks are located at ${\mathbf Q}=\pm$(0 1), $\pm$(0.5 0.5), and $\pm$(0.5 -0.5). (d) $\Psi$-dependence of the diffuse peak intensities for Sample$\#$1 (open symbol) and $\#2$ (closed symbol). Solid lines show the calculated $\Psi$-dependence for $x$-, $y$-, and $z$-zig-zag states shown in (e) for the $\pi$-$\sigma'$ and $\pi$-$\pi'$ polarization channels summed. (e) Zig-zag orders propagating along three equivalent directions. Blue zig-zag is the static structure, and green and yellow zig-zags are generated by 120$^\circ$ rotation of the blue zig-zag.
}\label{fig:fig2}
\end{figure*}

We start from establishing the spin orientation in the static zig-zag order\cite{Liu11,Choi12,Ye12} below T$_{\textrm N}$, as shown in Fig.~1a, using standard resonant magnetic x-ray diffraction. In this measurement, the x-ray polarization projects out a certain spin component; the intensity depends on the spin orientation via the relation $I\propto|{\mathbf k_f} \cdot {\mathbf S}|^2$ for the $\sigma$-$\pi'$ channel measured, where ${\mathbf k_f}$ is the scattered x-ray wave vector (Fig.~1b). Figure 1c shows the intensity variation as the sample is rotated about the ordering wave vector ${\mathbf Q}$ = (0 1 3.5) by an azimuth angle $\Psi$, which causes ${\mathbf S}$ to precess around ${\mathbf Q}$. Earlier studies\cite{Liu11,Ye12} have established that ${\mathbf S}$ is constrained to lie in the $ac$ plane, so this measurement of $I(\Psi)$ determines the spin orientation by resolving the tilting angle $\Theta$ of ${\mathbf S}$ with respect to the $a$ axis. The best fitting result with $\Theta$=44.3$^\circ$  indicates that the magnetic easy axis is approximately half way in between the cubic $x$ and $y$ axes (Fig.~1a). This static spin orientation is a compromise among all anisotropic interactions present in the system, and is strongly tied to the magnetic structure because of their bond-directional nature. To see this point, consider, for example, the $K$ term: in the zig-zag structure propagating along $b$ direction, where the spins are antiferromagnetically aligned on the $z$-bond and ferromagnetically aligned on the $x$-bond and $y$-bond, a ferromagnetic (antiferromagnetic) $K$ favors spins pointing perpendicular to (along) the $z$-axis for a pair of spins on the $z$-bond, and along (perpendicular to) the $x$-axis and $y$-axis for the pairs on the $x$-bond and $y$-bond, respectively.

The zig-zag order is one of the many magnetic states (including the aforementioned spiral and non-coplanar structures) that are classically degenerate in the pure Kitaev limit\cite{Price13} and comprise the micro-states in the QSL phase. Away from the pure Kitaev limit, depending on their energy separations, signatures of other magnetic states and their associated magnetic anisotropies may become observable in the paramagnetic phase through diffuse magnetic scattering. In particular, zig-zag orders propagating in two other directions, $\pm120^{\circ}$ rotated from the static one, are expected for a honeycomb net with $C_3$ symmetry. [The actual 3D crystal structure has only an approximate $C_3$ symmetry because of a monoclinic distortion, which singles out one propagation direction for the long-range ordered state (along $b$ direction) out of the three possible under the ideal $C_3$ symmetry\cite{Choi12}.]

With other magnetic correlations possibly emerging at high temperature in mind, we follow the temperature evolution of the zig-zag order. Figures 1d-f show $H$, $K$, and $L$ scans, respectively, of the magnetic Bragg peak at ${\mathbf Q}$ = (0 1 6.5) for selected temperatures. Figure 1g shows the correlation lengths along the $a$, $b$, and $c$ axes as a function of temperature. As the temperature increases above T$_{\textrm N}$, the zig-zag correlations diminish rather isotropically despite dominant 2D couplings in the honeycomb net. This 3D characteristic of the magnetic correlations contrasts with that of the quasi-2D Heisenberg antiferromagnet \SIOns, which displays 2D long-range correlations well above T$_{\textrm N}$\cite{Fujiyama12}, and implies that the critical temperature in \NIO is limited by the anisotropic interactions rather than the interlayer coupling; the Mermin-Wagner theorem requires either the symmetry to be lower than SU(2) or the dimension to be higher than 2D for a finite temperature phase transition. The zig-zag correlations survive on the length scale of several nanometers (approximately 3 unit cells wide) above T$_{\textrm N}$, but the peak intensities drop by two orders of magnitude. To isolate such small signals from the background, we used an experimental setup that maximizes the signal-to-noise ratio as described in the Methods section.

Figure 2a maps the diffuse scattering intensity over a region in momentum space encompassing a full Brillouin zone of the honeycomb net, at several different $\Psi$ angles to resolve the spin-components (see Fig.~S2). These maps integrate the dynamic structure factor over the range 0 $\leq$$ \omega$ $\leq$100 meV, covering the entire range of magnetic excitations (see Fig.~4), and serve as an excellent approximation for the equal-time correlation $S_{\alpha\alpha}\equiv\langle S^\alpha_{\mathbf Q} S^\alpha_{-\mathbf Q} \rangle$ ($\alpha=x,y,z$).   When averaged over the three spin components, the intensity map (Fig. 2c) indeed shows three zig-zag correlations above T$_{\textrm N}$ with peaks at ${\mathbf Q}$ = $\pm$(0 1), $\pm$(0.5 0.5), and $\pm$(0.5 -0.5) of equal intensities, confirming the near-ideal $C_3$ symmetry. However, the spin-component resolved maps, shown in Fig.~2b, manifestly break the $C_3$ symmetry. The system is left invariant only when $C_3$ rotation is performed simultaneously in the real space and in the spin space, i.e. cyclic permutation of spin indices. This `global' $C_3$ symmetry implies a strong entanglement between the real space and the spin space.  Specifically, the full azimuth dependence of each zig-zag state, shown in Fig.~2d, follows well the curves simulated for spin orientation fixed relative to the propagation direction, as depicted in Fig.~2e. In other words, specifying a spin component amounts to fixing the momentum direction and vice versa. This one-to-one correspondence between the spin space and the real space is a direct consequence of the bond-dependent nature of the anisotropic exchange terms. 

%
%

\begin{figure}[t]
\hspace*{-0.2cm}\vspace*{-0.1cm}\centerline{\includegraphics[width=1\columnwidth,angle=0]{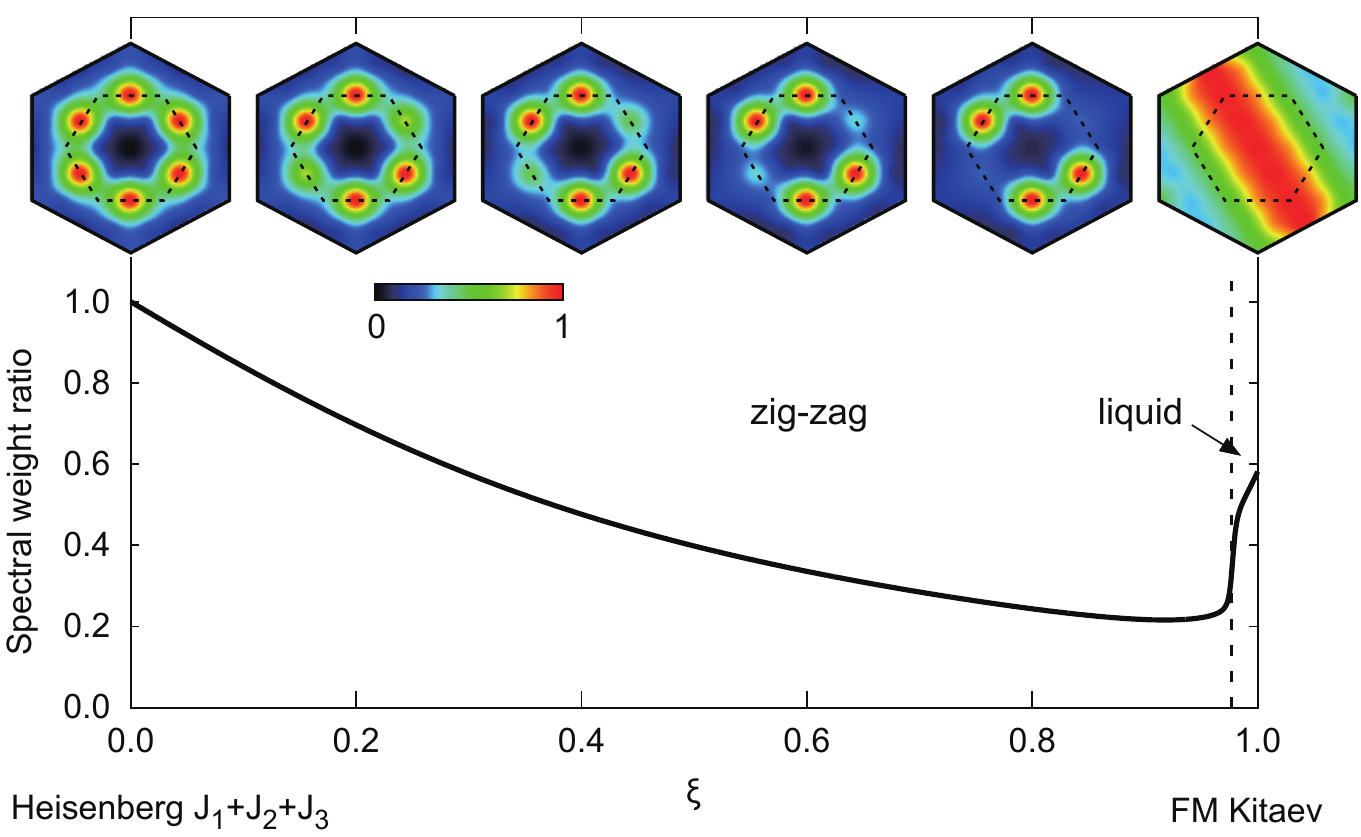}}

\caption{{\bf Simulation of diffuse scattering using exact diagonalization.} The Kitaev-Heisenberg model including up to third nearest-neighbor Heisenberg interactions was considered. $\xi$ interpolates between the pure Heisenberg model and the pure Kitaev model via $J_1$=$J_2$=$J_3$=$1- \xi$ and $K=-\xi$. A ferromagnetic $K$ with finite $J_1$,$J_2$, and $J_3$ stabilizes the zig-zag state for most values of $\xi$. The black curve shows the spectral asymmetry defined as the ratio of spectral weight at ${\mathbf Q}=$(0.5,0.5) to that at ${\mathbf Q}=$ (0 1). Images show equal-time correlations $\langle S^x_{\mathbf Q} S^x_{-\mathbf Q} \rangle$ obtained by exact diagonalization using a 24--cite cluster and plotted in the extended Brillouin zone for selected $\xi$. The correlations for $y$ and $z$ components (not shown) can be generated by $\pm$120$^\circ$ rotations of the images shown.
}\label{fig:fig3} 
\end{figure}

Qualitatively, it is immediately seen that the anisotropic interactions dominate over the isotropic interactions and the system is very far away from the pure Heisenberg limit, in which case the spatial correlations must be spin-component independent with three zig-zag peaks having equal intensities by symmetry (as in the spin-averaged correlation shown in Fig.~2c preserving $C_3$ symmetry). A measure of how close the system is to either the Heisenberg or the Kitaev limits is provided by the intensity ratio of the weakest peak to the two bright ones in the spin-component-resolved correlations (Fig.~2b). To quantify this measure, represented by a variable linearly interpolating between these two limits, $\xi$, requires specifying the minimal Hamiltonian, which is not precisely known. For an estimation at a semi-quantitative level, we adopt a simple Hamiltonian that neglects all anisotropic terms beyond the $K$ term. (This in turn requires including further neighbor Heisenberg couplings $J_2$ and $J_3$ to stabilize the zig-zag order\cite{Kimchi11}, which we take to be equal to $J_1$ for simplicity.) Figure 3 shows the simulated patterns for selected $\xi$. It is clear that the observed diffuse pattern is consistent with the simulated one for the large $\xi$ limit. In fact, the observed intensity ratio of $\approx$0.2 is even smaller than calculated (Fig.~3) for the largest $\xi$ in the zig-zag phase, which confirms the predominant anisotropic interactions.  

Interpreted within this model, our calculations would imply that the system is very close to the Kitaev limit. However, it is becoming increasingly evident that other anisotropic terms beyond the Kitaev interaction do play a role\cite{Rau14,Lee}. This is, in fact, evident from the static spin not pointing along one of the cubic axes favored by the $K$ term; all other anisotropic terms conspire to rotate the spin away from the principal axes. This in turn suggests that the zig-zag structure is further stabilized by other anisotropic terms. The zig-zag correlations survive at least up to $\sim$70 K (see Fig.~S3), which is in accord with the observation that coherent spin-waves\cite{Choi12} disperse up to $\approx$5 meV. This energy scale coincides with the temperature scale ($\approx$100 K) below which the magnetic susceptibility deviates from the Curie-Weiss behavior\cite{Singh}. This energy scale is, however, still far too small in comparison to the energy ($\approx$100 meV) spanned by the magnetic excitations (Fig. 4), suggesting that the zig-zag order is an emergent phenomenon. Despite that the macroscopic degeneracy in the Kitaev QSL phase is reduced down to three zig-zags, the high-energy Kitaev interactions leave their signature in the low-energy sector: the three spin components, each carrying its own zig-zag, compete and melt the long-range order at a temperature much lower than that suggested by the Weiss temperature ($\Theta_{\textrm W}$), leading to a large frustration parameter\cite{Singh} ($\equiv\Theta_{\textrm W}$/T$_{\textrm N}$) of $\approx$8.

\begin{figure}
\hspace*{-0.2cm}\vspace*{-0.1cm}\centerline{\includegraphics[width=0.9\columnwidth,angle=0]{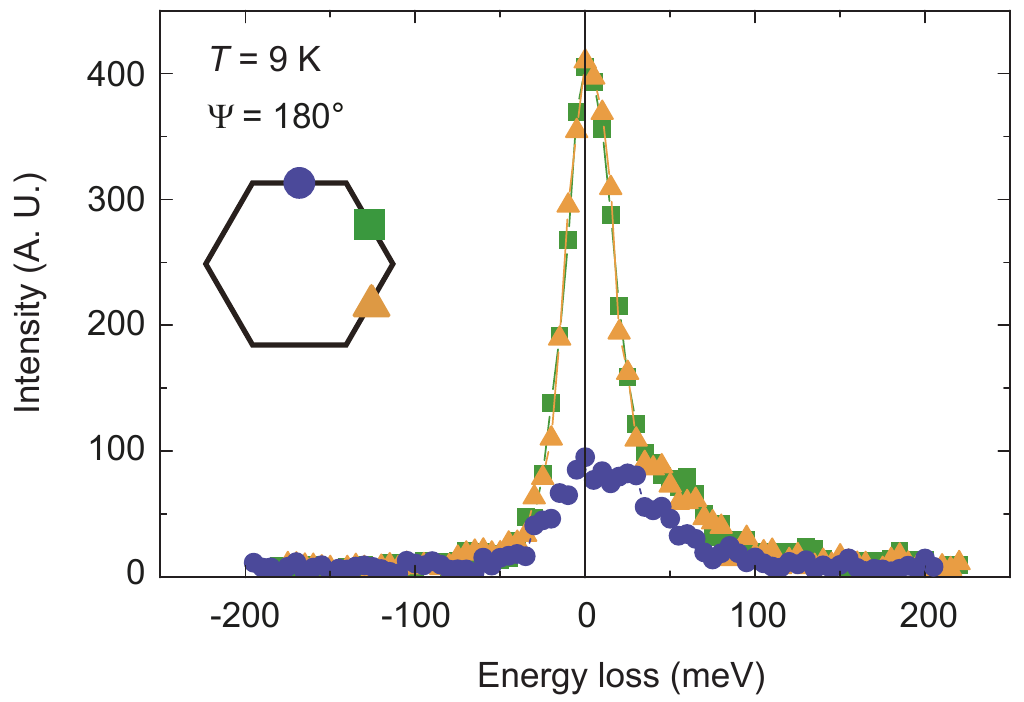}}

\caption{{\bf Resonant inelastic x-ray scattering spectra below T$_{\textrm N}$.} RIXS spectra recorded at T=9 K and $\Psi$=180$^\circ$. ${\mathbf Q}$ = (0 1), (0.5 0.5), and (0.5 -0.5), shown as blue, green, and yellow filled symbols, respectively, marked on the Brillouin zone of the honeycomb net and color-coded with the spectra. At this $\Psi$ angle, ${\mathbf S}$ lies approximately along ${\mathbf k_f}$ and $\pi$-$\sigma$' and $\pi$-$\pi$' channels measure the two spin components transverse to ${\mathbf S}$.
}  
\end{figure}

Beneath a rather mundane static collinear antiferromagnetic structure, the spins on the honeycomb lattice undergo fluctuations of a peculiar type that suppress the tendency to order. To reveal their full dynamics remains a challenge, but RIXS allows a glimpse into their unconventional nature.  Figure 4 shows the excitation spectra recorded in the ordered phase, in the scattering geometry that probes two spin components transverse to the static moment. The soft excitations ($\omega$$\approx$0) are observed at ${\mathbf Q}=\pm$(0.5 0.5) and ${\mathbf Q}=\pm$(0.5 -0.5), away from the Bragg peak at ${\mathbf Q}=\pm$(0 1). This is a notable exception to the universality held in conventional magnets that spin waves emanate from Bragg peaks by virtue of the Goldstone theorem, and magnetic anisotropy is manifested as a spin-wave gap, even in systems with extremely large magnetic anisotropy\cite{Kim12}. By contrast, the spin gap in our system is small (unresolved in our spectra and estimated to be smaller than 2 meV from INS data\cite{Choi12}) compared to the overall energy scale of the system, despite the fact that the magnetism is dominated by the anisotropic terms. Rather, the anisotropy is manifested as the separation of the long-wavelength spin waves from the Bragg peaks, which is a natural consequence of each spin component displaying its own real-space correlations. Our results directly reveal the key building blocks of the Kitaev model in \NIOns, and establishes a new design strategy for the long-sought quantum spin liquids via the bond-directional magnetic couplings.

\vspace{5mm}
\noindent\textbf{Acknowledgements} Work in the Materials Science Division of Argonne National Laboratory (sample preparation, characterization, and contributions to data analysis) was supported by the U.S. Department of Energy, Office of Science, Basic Energy Sciences, Materials Science and Engineering Division. Use of the Advanced Photon Source, an Office of Science User Facility operated for the U.S. Department of Energy (DOE) Office of Science by Argonne National Laboratory, was supported by the U.S. DOE under Contract No. DE-AC02-06CH11357. K.M. acknowledges support from UGC-CSIR, India. Y.S. acknowledges DST, India for support through Ramanujan Grant \#SR/S2/RJN-76/2010 and through DST grant \#SB/S2/CMP-001/2013. J.C. was supported by ERDF under project CEITEC (CZ.1.05/1.1.00/02.0068) and EC 7$^\mathrm{th}$ Framework Programme (286154/SYLICA).\\\\
\noindent\textbf{Author contribution} B.J.K. conceived the project. S.H.C., J.-W.K., J.K., and B.J.K. performed the experiment with support from Y.C., T.G., A.A., M.M.S., and M.K. H.Z. and K.M. grew the single crystals; C.S., C.D.M., and K.M. characterized the samples under supervision of J.F.M. and Y.S. S.H.C., J.-W.K., and B.J.K. analyzed the data. J.C. performed the numerical calculations; J.C., G.J., and G.K. developed the theoretical model. All authors discussed the results. B.J.K. led the manuscript preparation with contributions from all authors.\\\\
\noindent\textbf{Competing Interests} The authors declare that they have no competing financial interests.\\\\
\noindent\textbf{Correspondence} Correspondence and requests for materials should be addressed to  B.J.K.~(bjkim@fkf.mpg.de).\\\\

\newpage
\break
\noindent\textbf {Supplementary Information}\\\\
\noindent\textbf {A. Methods}\\\\
\noindent
\noindent\textbf {A.1 Single crystal growth \& characterization}\\
 Single crystals of \NIO were grown following two different recipes using Na$_2$CO$_3$ flux (Sample \#1) and self-flux (Sample \#2). For Sample \#1, a mixture of Na$_2$CO$_3$ and IrO$_2$ with a molar ratio of 50 : 1 was melted at 1050$^\circ$ C for 6 hours followed by fast cooling at a rate of 100$^\circ$ C/hour down to 1000$^\circ$ C, slow cooling at a rate of 1$^\circ$ C/hour down to 800$^\circ$ C and furnace-cooling to room temperature in a sequence. Hexagonal pillar-shaped crystals with typical dimension of 0.2 mm $\times$ 0.2 mm $\times$ 0.4 mm were obtained after dissolving Na$_2$CO$_3$ flux in acetone and water. For Sample\#2, powders of Na$_2$CO$_3$ were mixed with 10-20\% excess IrO$_2$ and were calcined at 700$^\circ$ C for 24 hours. Single crystals were grown on top of a powder matrix in a subsequent heating at 1050$^\circ$ C. Plate-like crystals with typical dimension of 5 mm $\times$ 5 mm $\times$ 0.1 mm were physically extracted.

\begin{figure}[b]
\setcounter{figure}{0}
\renewcommand{\thefigure}{S\arabic{figure}}
\hspace*{-0.2cm}\vspace*{-0.1cm}\centerline{\includegraphics[width=1\columnwidth,angle=0]{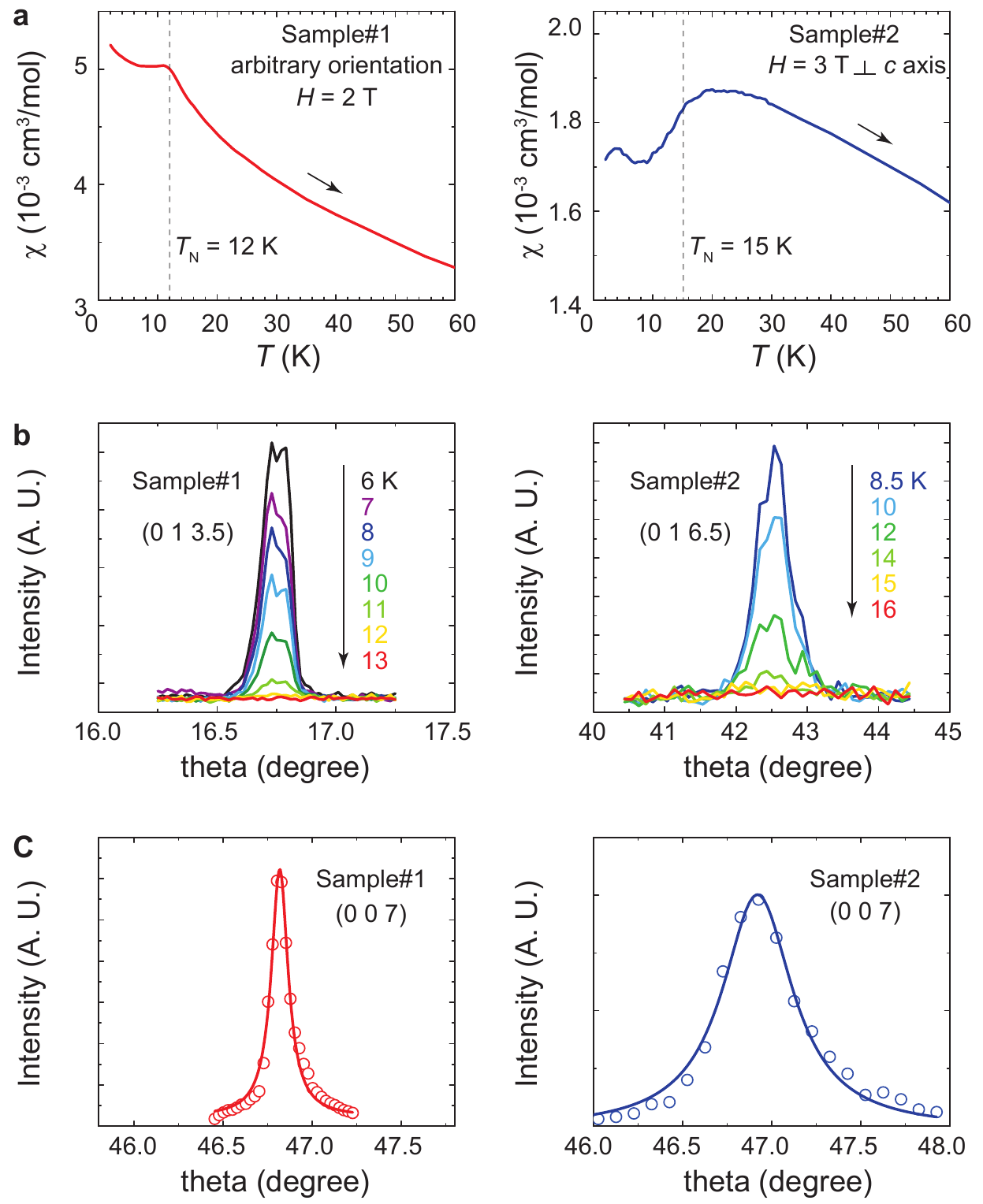}}

\caption{\textbf{Characterization of Sample \#1 and Sample \#2.} (a) Temperature dependence of magnetic susceptibility. Dotted lines indicate T$_{\textrm N}$. Black arrows indicate that the data were measured while warming after zero-field cooling. (b) Temperature dependence of the magnetic Bragg peaks. (c) Sample mosaicity.
}\label{fig:fig3} 
\end{figure}

 The powder x-ray diffraction patterns of both Sample \#1 and \#2 were consistent with the crystal structure in the C2/m space group as previous reported$^{15}$. Sample \#1 had a slightly lower T$_{\textrm N}$=12 K compared to Sample \#2 with T$_{\textrm N}$=15 K, as measured by SQUID magnetometry (Fig.~S1a) and by resonant x-ray diffraction through the magnetic Bragg peaks at Q = (0 1 n+$\frac{1}{2}$) (n: integer) (Fig.~S1b). Sample \#1 was found to be of multi domains but had a superior crystallinity with 0.1$^\circ$ mosaicity (as compared to 0.5$^\circ$ mosaicity of Sample \#2) (Fig.~S1c), and thus was used for the resonant diffraction experiment (polarization analysis and measurement of the magnetic correlation length.) Sample \#2 was found to be of a single domain and was used for the RIXS measurement. Both Sample \#1 and \#2 were used for the diffuse scattering measurement and gave identical results.

\vspace{5mm}
\noindent\textbf {A.2 Resonant x-ray scattering}\\
\noindent
Incident x-ray was tuned to Ir L$_3$ edge (11.2145 keV). The resonant x-ray diffraction was carried out at 6 ID-B beamline of the Advanced Photon Source. The polarization analysis was performed in the vertical scattering geometry using a pyrolytic graphite analyzer probing $\sigma$-$\pi'$ channel. The RIXS was performed at ID20 of the European Synchrotron Radiation Facility. The total instrumental energy resolution of 24 meV was achieved with a monochromator and a diced spherical analyzer made from Si (844) and a position-sensitive area detector placed on a Rowland circle with 2m radius. The diffuse magnetic scattering was performed using the RIXS spectrometers at 9 ID, 27 ID, and 30 ID (MERIX) beamlines of the Advanced Photon Source where a monochromator of 90 meV bandwidth was used for an order of magnitude higher incident photon flux than that from the Si (844) monochromator. In these experiments, horizontal scattering geometry was used with the $\pi$-incident x-ray polarization measuring the sum of $\pi$-$\sigma'$ and $\pi$-$\pi'$ channels. The in-plane momentum resolution of the RIXS spectrometer was $\pm$0.048 \AA$^{-1}$. The use of the RIXS spectrometers rejecting all inelastically scattered x-rays outside of the 100 meV energy window centered at the elastic line led to a significant improvement in the signal-to-noise 
ratio. A typical counting time of 2 hours was required for a map shown in Fig.~2.  

\vspace{5mm}
\noindent\textbf {B. Extraction of the spin-component resolved equal-time correlators}\\
\noindent
The x-ray scattering intensity measured without using a polarization analyzer contains contributions from both $\pi$-$\sigma'$ and $\pi$-$\pi'$ channels, probing spin components along ${\mathbf k_i}$ and perpendicular to the horizontal scattering plane, respectively. In other words, two spin components perpendicular to ${\mathbf k_f}$ are measured in the 90$^\circ$ horizontal scattering geometry used. For example, when $\Psi$=0$^\circ$, the $z$ local cubic axis  points approximately along ${\mathbf k_f}$ (Fig.~S2), and thus the scattering intensity measures the correlation $S_{xx}$+$S_{yy}$. Likewise, $\Psi$=120$^\circ$($\Psi$=240$^\circ$) measures $S_{xx}$+$S_{zz}$($S_{yy}$+$S_{zz}$). Then, $S_{xx}$, $S_{yy}$, and $S_{zz}$ can be extracted by solving a set of linear equations.

\begin{figure}[h]
	\renewcommand{\thefigure}{S\arabic{figure}}
	\hspace*{-0.2cm}\vspace*{-0.1cm}\centerline{\includegraphics[width=1\columnwidth,angle=0]{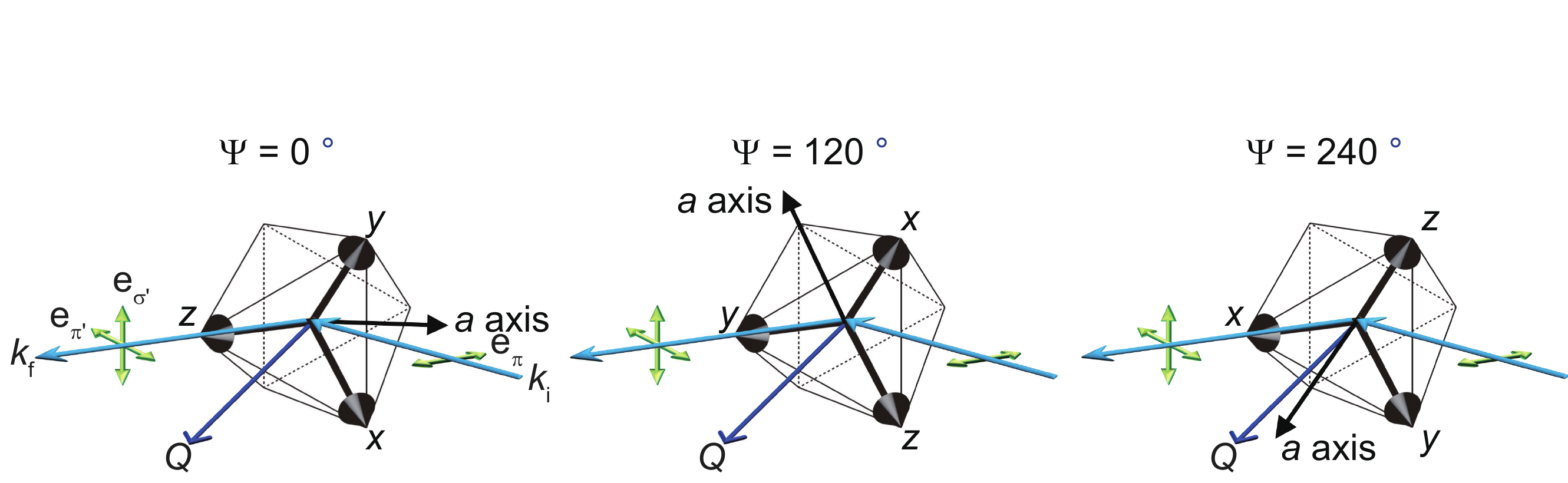}}
	
	\caption{\textbf{Scattering configurations.} (a) $\Psi$= 0$^\circ$, (b) $\Psi$ = 120$^\circ$, and (c) $\Psi$ = 240$^\circ$. The azimuth $\Psi$ is the angle between the $a$ axis and the scattering plane. Thick black arrows denote the local $x$, $y$, and $z$ axes in the IrO$_6$ octahedron. $a$ axis is indicated by a thin black arrow. Cyan arrows are the incident and scattered x-rays with the wave vectors ${\mathbf k_i}$ and ${\mathbf k_f}$, respectively. Green arrows indicate the x-ray polarizations. ${\mathbf Q}$ (dark blue arrow) is the momentum transfer. 
	}\label{fig:fig3} 
\end{figure}

\vspace{5mm}
\noindent\textbf {C. Temperature dependence of the diffuse magnetic peak}\\
\noindent
The short-range zig-zag order is observable at least up to T$\approx$70 K (Fig.~S3a). The magnetic correlation length (1.6-1.8 nm) along the $a$ axis does not vary significantly in the measured temperature region. Fig.~S3b plots the diffuse map at $\Psi$=0$^\circ$ at T=50 K, which is similar to the diffuse map recorded at T=17 K (shown in Fig.~2a) apart from thermal broadenings. \\\\

\begin{figure}[h]
\renewcommand{\thefigure}{S\arabic{figure}}
\hspace*{-0.2cm}\vspace*{-0.1cm}\centerline{\includegraphics[width=1\columnwidth,angle=0]{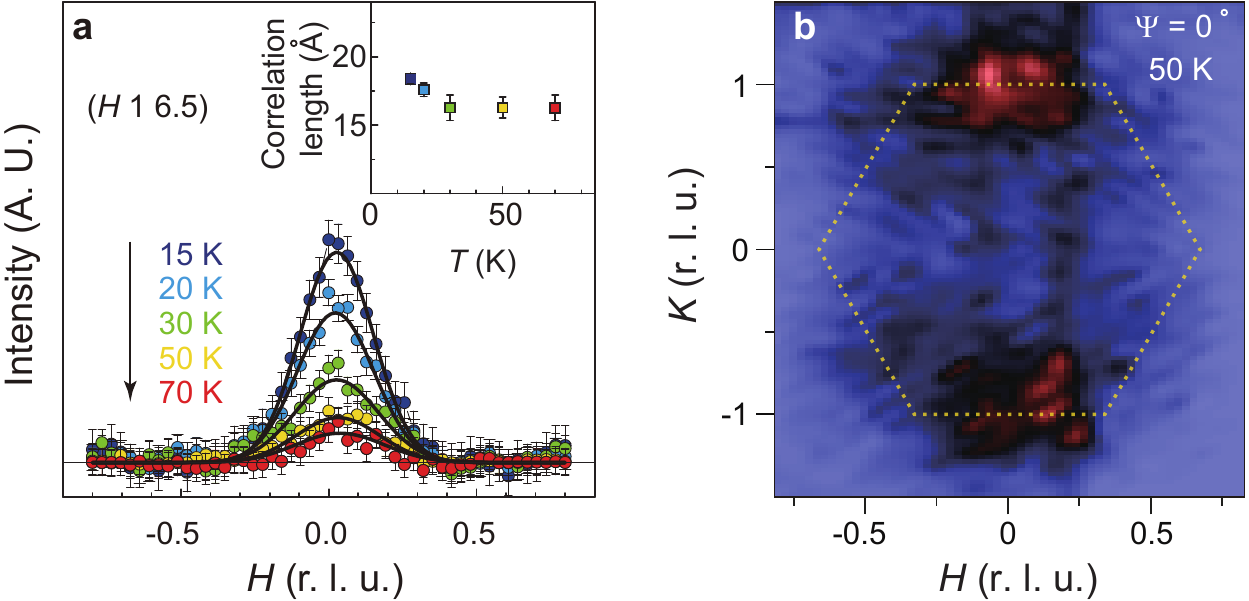}}

\caption{\textbf{Diffuse magnetic x-ray scattering intensities above T$_{\textrm N}$.} (a) H profiles of the diffuse peak at Q = (0 1). Shown in the inset is the magnetic correlation length along the $a$ axis derived from Gaussian fitting (solid curves)of the data for Sample \#1. (b) Diffuse map at $\Psi$=0$^\circ$ at T=50 K. The dashed hexagon indicates the first Brillouin zone of the honeycomb net.
}
\end{figure}


\begin{thebibliography}{99}

\subsection*{References}
\bibitem{vanVleck}	van Vleck, J. H. On the anisotropy of cubic ferromagnetic crystals. \textit{Phys. Rev.} \textbf{52}, 1178 (1937).
\bibitem{Khaliullin05} Khaliullin, G. Orbital Order and Fluctuations in Mott Insulators. \textit{Prog. Theor. Phys. Suppl.} \textbf{160}, 155 (2005).
\bibitem{Jackeli09}	Jackeli, G. \& Khaliullin, G. Mott insulators in the strong spin-orbit coupling limit: from Heisenberg to a quantum compass and Kitaev models. \textit{Phys. Rev. Lett.} \textbf{102}, 017205 (2009). 
\bibitem{Kitaev}	Kitaev, A. Anyons in an exactly solved model and beyond. \textit{Ann. Phys.} \textbf{321}, 2 (2006).

\bibitem{Kim08}	Kim, B. J. \textit{et al.} Novel $J_{\textrm eff}$ = 1/2 Mott state induced by relativistic spin-orbit coupling in Sr$_2$IrO$_4$. \textit{Phys. Rev. Lett.} \textbf{101}, 076402 (2008).
\bibitem{Kim09}	Kim, B. J. \textit{et al.} Phase-sensitive observation of a spin-orbital Mott state in \SIOns. \textit{Science} \textbf{323}, 1329 (2009).
\bibitem{Takayama}	Takayama, T. \textit{et al.} Hyper-honeycomb iridate $\beta$-Li$_2$IrO$_3$ as a platform for Kitaev magnetism. \textit{Phys. Rev. Lett.} \textbf{114}, 077202 (2015).
\bibitem{Modic}	Modic, K. A. \textit{et al.} Realization of a three-dimensional spin-anisotropic harmonic honeycomb iridate. \textit{Nat. Commun.} \textbf{5}, 4203 (2014).
\bibitem{Kimchi14}	Kimchi, I. \& Vishwanath, A. Kitaev-Heisenberg models for iridates on the triangular, hyperkagome, kagome, fcc, and pyrochlore lattices. \textit{Phys. Rev. B} \textbf{89}, 014414 (2014).
\bibitem{Hermanns}	Hermanns, M. \& Trebst, S. Quantum spin liquid with a Majorana Fermi surface on the three-dimensional hyperoctagon lattice. \textit{Phys. Rev. B} \textbf{89}, 235102 (2014).
\bibitem{Lee14}	Lee, S. B., Jeong, J.-S., Hwang, K. \& Kim, Y. B. Emergent quantum phases in a frustrated $J_1$-$J_2$ Heisenberg model on the hyperhoneycomb lattice. \textit{Phys. Rev. B} \textbf{90}, 134425 (2014).
\bibitem{Plumb14}	Plumb, K. W. \textit{et al.}, $\alpha$-RuCl$_3$: a spin-orbit assisted Mott insulator on a honeycomb lattice. \textit{Phys. Rev. B} \textbf{90}, 041112 (2014).
\bibitem{Luo13}	Luo, Y. \textit{et al.}, Li$_2$RhO$_3$: A spin-glassy relativistic Mott insulator. \textit{Phys. Rev. B} \textbf{87}, 161121(R)  (2013).
\bibitem{Liu11}	Liu, X. \textit{et al.} Long-range magnetic ordering in \NIOns. \textit{Phys. Rev. B} \textbf{83}, 220403(R) (2011).
\bibitem{Choi12}	Choi, S. K. \textit{et al.} Spin waves and revised crystal structure of honeycomb iridate \NIOns. \textit{Phys. Rev. Lett.} \textbf{108}, 127204 (2012).
\bibitem{Ye12}	Ye, F. \textit{et al.} Direct evidence of a zigzag spin-chain structure in the honeycomb lattice: A neutron and x-ray diffraction investigation of single-crystal \NIOns. \textit{Phys. Rev. B} \textbf{85}, 180403(R) (2012).
\bibitem{Reuther14}	Reuther, J., Thomale, R. \& Rachel, S. Spiral order in the honeycomb iridate Li$_2$IrO$_3$.  \textit{Phys. Rev. B} \textbf{90}, 100405(R) (2014).
\bibitem{Biffin14}	Biffin, A. \textit{et al.} Noncoplanar and counterrotating incommensurate magnetic order stabilized by Kitaev Interactions in $\gamma$-Li$_2$IrO$_3$. \textit{Phys. Rev. Lett.} \textbf{113}, 197201 (2014).
\bibitem{Biffin142}	Biffin, A. \textit{et al.} Unconventional magnetic order on the hyperhoneycomb Kitaev lattice in $\beta$-Li$_2$IrO$_3$: Full solution via magnetic resonant x-ray diffraction. \textit{Phys. Rev. B} \textbf{90}, 205116 (2014).
\bibitem{Chaloupka10}	Chaloupka, J., Jackeli, G. \& Khaliullin, G. Kitaev-Heisenberg model on a honeycomb lattice: possible exotic phases in iridium oxides $A_2$IrO$_3$. \textit{Phys. Rev. Lett.} \textbf{105}, 027204 (2010). 
\bibitem{Chaloupka13}	Chaloupka, J., Jackeli, G. \& Khaliullin, G. Zigzag magnetic order in the iridium oxide \NIOns. \textit{Phys. Rev. Lett.} \textbf{110}, 097204 (2013).

\bibitem{Lee}	Lee, E. K.-H. \& Kim, Y. B. Theory of magnetic phase diagrams in hyperhoneycomb and harmonic-honeycomb iridates. \textit{Phys. Rev. B} \textbf{91} 064407 (2015).
\bibitem{Rau14}	Rau, J. G., Lee, E. K.-H. \& Kee, H.-Y. Generic spin model for the honeycomb iridates beyond the Kitaev limit. \textit{Phys. Rev. Lett.} \textbf{112}, 077204 (2014).
\bibitem{Katukuri14}	Katukuri, V. M. \textit{et al.} Kitaev interactions between $j$ = 1/2 moments in honeycomb \NIO are large and ferromagnetic: insights from ab initio quantum chemistry calculations. \textit{New J. Phys.} \textbf{16}, 013056 (2014).
\bibitem{Yamaji14}	Yamaji, Y., Nomura, Y., Kurita, M., Arita, R. \& Imada, M. First-principles study of the honeycomb-lattice iridates \NIO in the presence of strong spin-orbit interaction and electron correlations. \textit{Phys. Rev. Lett.} \textbf{113}, 107201 (2014).
\bibitem{Shitade09}	Shitade, A. \textit{et al.} Quantum spin Hall effect in a transition metal oxide \NIOns. \textit{Phys. Rev. Lett.} \textbf{102}, 256403 (2009).
\bibitem{Mazin12}	Mazin, I. I., Jeschke, H. O., Foyevtsova, K., Valenti, R. \& Khomskii, D. I. \NIO as a molecular orbital crystal. \textit{Phys. Rev. Lett.} \textbf{109}, 197201 (2012).
\bibitem{KimC12}	Kim, C. H., Kim, H. S., Jeong, H., Jin, H. \& Yu, J. Topological quantum phase transition in 5d transition metal oxide \NIOns. \textit{Phys. Rev. Lett.} \textbf{108}, 106401 (2012).
\bibitem{Gretarsson13}	Gretarsson, H.\textit{ et al.} Magnetic excitation spectrum of \NIO probed with resonant inelastic x-ray scattering. \textit{Phys. Rev. B} \textbf{87}, 220407 (2013).
\bibitem{Price13}	Price, C. \& Perkins, N. B. Finite-temperature phase diagram of the classical Kitaev-Heisenberg model. \textit{Phys. Rev. B} \textbf{88}, 024410 (2013).
\bibitem{Fujiyama12}	Fujiyama, S. \textit{et al.} Two-dimensional Heisenberg behavior of $J_{\textrm eff}$ = 1/2 isospins in the paramagnetic state of the spin-orbital Mott insulator \SIOns. Phys. Rev. Lett. 108, 247212 (2012).
\bibitem{Kimchi11}	Kimchi, I. \& You, Y.-Z. Kitaev-Heisenberg-$J_2$-$J_3$ model for the iridates $A_2$IrO$_3$. \textit{Phys. Rev. B} \textbf{84}, 180407(R) (2011).
\bibitem{Singh}	Singh, Y. \& Gegenwart, P. Antiferromagnetic Mott insulating state in single crystals of the honeycomb lattice material \NIOns. \textit{Phys. Rev. B} \textbf{82}, 064412 (2010).
\bibitem{Kim12}	Kim, J. \textit{et al.} Large spin-wave energy gap in the bilayer iridate Sr$_3$Ir$_2$O$_7$: Evidence for enhanced dipolar interactions near the Mott metal-insulator transition. \textit{Phys. Rev. Lett.} \textbf{109}, 157402 (2012).
\end{thebibliography}
\end{document}